\documentclass[10pt,a4paper,onecolumn]{article}

\usepackage[margin=3cm]{geometry} 
\usepackage{setspace}
\usepackage{titlesec}
\usepackage{enumitem}
\usepackage{lmodern}
\usepackage[T1]{fontenc}
\usepackage{fontspec} 
\usepackage{graphicx}
\usepackage{tcolorbox}
\usepackage{tipa}
\usepackage[utf8]{inputenc}
\usepackage{hyperref}
\usepackage[
    backend=biber,
    style=numeric-comp,
    language=english,
    sorting=none,
    date=long      
]{biblatex}

\usepackage{lettrine}
\newcommand{\parbreak}{\vspace{1em}}
\titleformat{\section}{\normalsize\bfseries}{\thesection.}{0.4em}{}
\titlespacing*{\section}{0pt}{1em}{0.5em}

\usepackage{setspace}
\usepackage{changepage}
\newenvironment{preambleblock}{
  \begin{center}
    \vspace{-1.5em}
    \rule{0.6\textwidth}{0.4pt}\\[0.8em]
    \begin{minipage}{0.9\textwidth}
    \begin{spacing}{1.15}
    \itshape
}{
    \end{spacing}
    \end{minipage}\\[0.8em]
    \rule{0.6\textwidth}{0.4pt}
  \end{center}
  \vspace{1.5em}
}

\newenvironment{signatures}{
  \section*{\large Signatories}
  \footnotesize
  \begin{itemize}[leftmargin=*, label={}, itemsep=0.1em]
}{
  \end{itemize}
}
\title{\vspace{-2.0em}
\textbf{Quantum Scientists for Disarmament: a Manifesto}}
\author{Quantum Scientists for Disarmament$^{*,\dagger}$}
\date{}

\begin{document}
\maketitle
\renewcommand{\thefootnote}{\fnsymbol{footnote}} 

\footnotetext[1]{All members of this collaboration are listed with their affiliations as signatories at the end of the document.}
\footnotetext[2]{Corresponding authors: Marco Cattaneo (\href{mailto:marco.cattaneo@helsinki.fi}{marco.cattaneo@helsinki.fi}), Luca Tagliacozzo (\href{mailto:luca.tagliacozzo@iff.csic.es}{luca.tagliacozzo@iff.csic.es}), Raja Yehia (\href{mailto:raja.yehia@icfo.eu}{raja.yehia@icfo.eu}).}

\renewcommand{\thefootnote}{\arabic{footnote}} 
\date{}

\vspace{-2.0em} 

\begin{preambleblock}
\vspace{1.0em}
We, as researchers in quantum science and technology, are publishing this manifesto to express our deep concerns about the current geopolitical situation and the global race to rearm. We firmly oppose all forms of militarization in our societies and, in particular, within the academic world. We categorically reject the use of our research for military applications, population control, or surveillance. We stand against the practice of military funding for research. 
This manifesto is a call to action: to confront the elephant in the room of quantum research, and to unite all researchers who share our views. 

Our main goals are:
\begin{itemize}[leftmargin=1em, itemsep=0.2em]
    \item To express, as a unified collective, our \textbf{rejection of the use of our research for military purposes}.
    \item To \textbf{open a debate} in our community about the ethical implications of quantum research for military purposes.
    \item To \textbf{create a forum} where concerned scientists can share their opinions and join forces in support of demilitarized research. 
    \item To advocate for the \textbf{establishment of a public database} listing all research projects at public universities funded by military or defense agencies.
\end{itemize}

In what follows, we lay out our concerns and the rationale behind our opposition to the militarization of quantum research.
\end{preambleblock}

War is once again raging across the globe, with estimates ranging from 70 to more than 110 armed conflicts currently ongoing \cite{geneva2025,crisis2025}.
In this already worrying context, we are witnessing an accelerated global race toward rearmament \cite{sipri2025fact,sipri2025press,undp2025,iiss2025}. While both East and West Asia are also a key part of this global trend, the arms race is escalating most dramatically in the European continent. Russia has recently undergone a drastic process of militarization \cite{sipri2025fact,sipri2025press}. Ukraine, in the middle of an excruciating war, currently spends more than 30\% of its GDP on defense \cite{sipri2025fact,sipri2025press}. Moreover, mostly in response to the Russian invasion of Ukraine and under Trump’s influence, many countries of the European Union (EU) have started an unprecedented race to rearm \cite{santopinto2025}. 

\parbreak

The total defense budget of the EU member states is currently the second highest in the world, right after the USA \cite{consilium2025,sipri2025milex,defensepost2025,eurasiantimes2025}. Despite this fact, all NATO countries have committed to increasing their national defense budgets to as much as 5\% of their GDP \cite{reuters2025nato}. These countries have already hit the 2\% target in 2025 \cite{reuters2025allnato}, including those that are considered to have limited room for public spending, such as Italy \cite{reuters2025italy}.

\parbreak

The research and development sector is not exempt from this trend. On the contrary, it is for example explicitly identified as one of the main sectors where national governments of EU member states may choose to allocate the funds provided by the ReArm Europe Plan (currently known as the “Readiness 2030 Plan”) \cite{europarl2025}\textcolor{black}{, while the expansion of research projects focused on dual-use technologies had already been advocated in a 2024 white paper \cite{euDualUse2024}}.  As researchers in quantum information and quantum technologies, we are also witnessing a growing number of military-oriented research projects and applications in our fields \cite{krelina2021quantum,atarc2025,spie2025}, both in the private and public sectors. Realistic applications include, for example, quantum key distribution and cryptographic networks for communication between military forces, space quantum radars for satellite surveillance, quantum sensing and clocks for military navigation and positioning, and quantum sensors for drones \cite{krelina2021quantum}.

\parbreak 

We acknowledge that new technologies, including quantum technologies, are not neutral. For example, machine learning and cloud computing are already being used to repress populations by processing datasets of individuals and manipulating opinions \cite{carnegie2019}. Likewise, quantum technologies can enhance many of the tools employed in the race to rearm, as recognized by different national defense departments and international alliances \cite{nato2025topics,nato2025texts,eu2025quantum,quantumzeitgeist2025,itif2024,gov2020,quantuminsider2025,dod2025}. For instance, NATO considers quantum technologies part of the broader category of emerging and disruptive technologies (EDTs) that are critical to defense and security \cite{nato2025topics}. Moreover, in January 2024 NATO published its official Quantum Technology Strategy, aimed at building a “Quantum-ready Alliance” \cite{nato2025texts}. The European Union also clearly states that one of the objectives of developing quantum technologies in Europe is defense and security \cite{eu2025quantum}. Rosatom State Corporation, the Russian state organization for nuclear energy—including nuclear weapons—has also developed a quantum computing roadmap \cite{quantumzeitgeist2025}. China’s 14th five-year plan includes the strategic development of quantum technologies \cite{itif2024,gov2020}. \textcolor{black}{Moreover,} the “RDT\&E budget plan” of the U.S. Department of Defense (currently Department of War) includes consistent funding for quantum applications  \cite{quantuminsider2025,dod2025}. In the 2024 estimates for the 2025 budget, it was stated that the development of quantum technologies is “critical to maintaining the Nation’s technological superiority”, with the aim of creating a “quantum supply chain” that “will enable defense applications of quantum technology as well as dual-use applications” \cite{dod2024}. \textcolor{black}{In addition, India has launched the National Quantum Mission in strong collaboration with the public and private defense sector \cite{india}. Similar initiatives are emerging in several other countries, including for example Japan \cite{japan}, South Korea \cite{korea}, Iran \cite{iran}, Turkey \cite{turkey}, Brazil \cite{brazil}, and Peru \cite{peru}.}

\parbreak

The expansion of military funding for both basic and applied research on emerging technologies, including quantum technologies, is not limited to the world’s major military powers. In a broader context, this opaque expansion often takes the form of asymmetric military-academic partnerships between the defense departments of powerful nations and academic institutions of the Global South. This strategy serves as a subtle mechanism through which hegemonic countries impose their “soft” power over nations of the Global South. For instance, from the perspective of states that can spend less of their public money on science, these funds can support projects that would not be executed otherwise, and help maintain pre-existing infrastructure and personnel, appearing as nearly irrecusable offers \cite{fapesp2025}. The other facet of such military-academic agreements, devised by hegemonic powers, is the use of such funds as a way to make their national army “more lethal than ever”, as stated in the U.S. Army Combat Capabilities Development Command (DEVCOM) website \cite{devcom2025}.

\parbreak 

We, \textcolor{black}{as scientists whose research can contribute to the development of these new technologies}, are concerned about the current geopolitical situation and the response of all those governments around the world that are increasingly focusing on rearmament and expanding their military presence and influence within our societies. Historically, a race to rearm combined with the rise of far-right and nationalist movements---as we have been recently witnessing in many countries all over the world---has always led to major conflicts and to the loss of freedom and democracy. Weapons have never been built not to be used.

\parbreak

We are also extremely alarmed by declarations such as those made by the rector of one of the most important institutions in Germany, who claimed that universities should renounce their neutrality and start expanding the research lines focused on military applications, in the name of the defense of “freedom and democracy”, and criticized those universities that see themselves as “ivory towers” that do not want to work for the “needs of their country” \cite{spiegel2023}. These arguments completely ignore the rise of far-right movements across Europe, including in Germany: no one can guarantee that the military research conducted by public universities in so-called “liberal democracies” will not be used for illiberal purposes in the future. Illiberal purposes that may well be presented like the “needs of the country”, as it already happened in our darkest past. Once a dual-use technology is developed, it may be employed in the name of defending “freedom”, but just as easily for the needs of authoritarian regimes. 

\parbreak

Beyond this, we strongly believe that the \textcolor{black}{military} neutrality of universities and public research institutions is essential to ensure that public research serves the good of all humankind, rather than the geopolitical agenda of the local governments. The moment research institutions align themselves with the military goals of their countries, they surrender the very \textcolor{black}{independence} that gives their research legitimacy. Invoking the argument that we are in a moment of emergency, and that this \textcolor{black}{independence} may be reestablished once the emergency disappears, is both tricky and dangerous: once a university gets deeply entangled with defense departments, agencies, and the broader military-industrial complex—for instance by heavily relying on funding from these sources, as it happens in the USA \cite{gao2023,economicstrategy2025}—it becomes exceedingly difficult to reverse this trend. Doing so would require a radical restructuring of the university‘s financial system, possible only through strong and sustained political action.

\parbreak

Let us also consider that the argument that war is sometimes inevitable, commonly raised to advocate for a massive rearmament, sounds too often as a self-fulfilling prophecy rather than a well-grounded view of the current geopolitical situation, in which it is definitely not too late, for instance, to avoid a major war involving all European countries. A prophecy that is not only dangerous, but also instrumental in legitimizing an unprecedented shift of public money to the military sphere, and nurturing the seeds of militarism in different sectors of society, including the academic world. The history of the 20th century shows that the doctrine of peace through strength, which is still present in today’s political rhetoric \cite{ansa2025,pbs2025}, has too often led to devastating conflicts. Yet it continues to be invoked as a sophistic argument to reassure the European populations that the European rearmament is a guarantee of peace rather than a prelude to war.

\parbreak

In this evolving scenario, we would like to gain a clearer picture of how deeply the military has infiltrated the public landscape of quantum research in the world—and, more specifically, how this process has been affected by the recent massive rearmament in Europe. We want to understand whether and which groups working on quantum technologies have already begun performing classified research within public universities and research institutions, and/or receive research funding from defense agencies and the military-industrial complex. While partial information may already be available, it is often hidden and difficult to reach. For this reason, we advocate for the creation of a transparency database listing all projects in public universities that are funded by military or defense agencies, including those aimed at the development of dual-use technologies.
\parbreak

As an example of the ethical concerns such projects raise, Leonardo S.p.A.—Italy’s largest defense contractor, partly state-owned—participates in numerous dual-use research projects at public universities \cite{ilmanifesto2025}. At the same time, it collaborates with Israeli research institutions \cite{leonardo2023} and has a documented history of arms exports to Israel (whether and to which extent such transfers continue today remains uncertain and widely debated \cite{altreconomia2025}), \textcolor{black}{as well as to Egypt under Abdel Fattah el-Sisi’s regime \cite{LeonardoEgypt} and Saudi Arabia \cite{LeonardoSaudi}}. This means that\textcolor{black}{, among several other ethical concerns,} the development of dual-use technologies in partnership with Leonardo at public universities may have ultimately benefited the offensive military capability of the Israeli government, which a recent UN commission found responsible for genocide in Gaza \cite{ohchr2025}.

\parbreak

Against this background, we wish to distance ourselves from military-oriented applications of quantum technologies. We want to make sure that our findings are not used on the battlefield or as a means of repression. We wish to be part of a community of researchers more attentive to ethical issues and less focused on military-oriented applications or on profit-oriented projects that ignore ethical considerations. 

\parbreak

We also wish to foster open debate on these issues, create a network of concerned scientists, and establish a forum where we can express our opinions, organize campaigns, and join forces to feel less isolated in an increasingly militarized world. We would like to emphasize that in this debate, we also welcome the participation of scientists whose research is supported by military funding. Our intention is not to target individual behaviors, but rather to shed light on, critically examine, and ultimately seek to change the broader system of military involvement in academia. We also acknowledge that, in many contexts, researchers have very limited options when it comes to securing funding.

\parbreak

To conclude, we still believe that war must be utterly rejected as a means of settling international disputes, and that peace can only be guaranteed by diplomacy, international treaties, and cooperation, rather than by mutual assured destruction. As scientists working in a non-neutral research field, we can raise our voices toward that aim.

\vspace{2.0em}
\hfill January 2026

\vspace{1.0em}
\begin{signatures}
\item Nair Aucar Boidi — Abdus Salam International Centre for Theoretical Physics - ICTP, Italy
\item Antoine Baron — CNRS - Institut Néel, France, and Karlsruhe Institute of Technology, Germany
\item Paolo Battistoni — Karlsruhe Institute of Technology, Germany
\item Aleix Bou Comas — CSIC, Spain
\item Natalia Bruno —  CNR - Istituto Nazionale di Ottica (CNR-INO), LENS -  European Laboratory for Non-Linear Spectroscopy, Italy
\item Giacomo Carrara — Ghent University - IMEC, Belgium
\item Marco Cattaneo — University of Helsinki, Finland
\item Cecilia Chiaracane — University of Helsinki, Finland
\item Marilù Chiofalo — University of Pisa, Italy
\item Luca Chirolli — University of Florence, Italy
\item Danial Chughtai  — Technical University Vienna, Austria
\item Guilherme Ilário Correr — University of Helsinki, Finland
\item Flavio Del Santo  — University of Vienna, Austria
\item Emanuele Distante — University of Florence, Italy
\item Patrick Dreger Andriolo — Technische Universität Wien, Austria
\item Tomás Fernández Martos  — ICFO -  Institut de Ciencies Fotoniques, Spain
\item Caterina Foti  — Aalto University, Finland
\item Felipe Gewers  — Humboldt Universität zu Berlin, Germany
\item Gian Luca Giorgi  — Institute for Cross Disciplinary Physics and Complex Systems IFISC (CSIC - UIB), Spain
\item Paul Hilaire  — Télécom Paris, France
\item Marcus Huber  — Technical University Vienna, Austria
\item Fernando Iemini  — Universidade Federal Fluminense (UFF), Brazil
\item Nathan Keenan  — Institute for Cross Disciplinary Physics and Complex Systems IFISC (CSIC - UIB), Spain
\item Naga Bhavya Teja Kothakonda — Autonomous University of Barcelona (UAB), Spain
\item Esperanza Lopez — CSIC, Spain
\item Cosmo Lupo — Politecnico di Bari, Italy
\item Maria Maffei — University of Bari, Italy
\item Gonzalo Manzano  — Institute for Cross Disciplinary Physics and Complex Systems IFISC (CSIC - UIB), Spain
\item Pietro Massignan  — Universitat Politècnica de Catalunya, Spain
\item Raúl Morral-Yepes  — Technical University of Munich, Germany 
\item Pere Munar-Vallespir  —  Technical University of Munich, Germany
\item Manabendra Nath Bera  —  Indian Institute of Science Education and Research, Mohali, India
\item Simon Neves  —   Université Marie et Louis Pasteur, France
\item Gian Luca Oppo  —   University of Strathclyde, United Kingdom
\item Paulo Jose Paulino de Souza  — Institute for Cross Disciplinary Physics and Complex Systems IFISC (CSIC - UIB), Spain
\item Martí Perarnau-Llobet — Autonomous University of Barcelona (UAB), Spain
\item Francesco Plastina — University of Calabria, Italy
\item Nicola Pranzini — University of Helsinki, Finland
\item Leonardo Rincón Celis — Laboratoire Kastler Brossel, France
\item Alberto Rolandi  — Technical University of Vienna, Austria
\item Carlo Rovelli  —  Aix-Marseille University, France
\item Nahual Sobrino — Abdus Salam International Centre for Theoretical Physics - ICTP, Italy
\item Luca Tagliacozzo — Instituto de Física Fundamental - CSIC, Spain
\item Luisa Toledo Tude — Institute for Cross Disciplinary Physics and Complex Systems IFISC (CSIC - UIB), Spain
\item V. Vilasini — Inria, Université Grenoble Alpes, France
\item Otto Veltheim — University of Helsinki, Finland
\item Paola Verrucchi —  Istituto dei Sistemi Complessi - CNR \&  INFN \& Università di Firenze, Italy
\item Francesca Vidotto — Instituto de Estructura de la Materia (IEM-CSIC), Spain
\item Ludmila Viotti — Abdus Salam International Centre for Theoretical Physics - ICTP, Italy
\item Raja Yehia — ICFO -  Institut de Ciencies Fotoniques, Spain
\item Roberta Zambrini — Institute for Cross Disciplinary Physics and Complex Systems IFISC (CSIC - UIB), Spain

\end{signatures}

\vspace{1.0em}

\printbibliography[title={\large References}]

@online{geneva2025,
  title        = {Today's Armed Conflicts},
  organization = {Geneva Academy of International Humanitarian Law and Human Rights},
  url          = {https://geneva-academy.ch/galleries/today-s-armed-conflicts},
    urldate = {2026-01-07}
}

@online{crisis2025,
  title        = {CrisisWatch},
  organization = {International Crisis Group},
  year         = {2025},
  url          = {https://www.crisisgroup.org/crisiswatch},
    urldate = {2026-01-07}
}

@online{india,
  title        = {National Quantum Mission},
  organization = {Government of India},
  year         = {2023},
  url          = {https://dst.gov.in/national-quantum-mission-nqm},
    urldate = {2026-01-07}
}

@online{japan,
  title        = {Quantum Technology Innovation},
  organization = {Cabinet Office, Government of Japan},
  year         = {2025},
  url          = {https://www8.cao.go.jp/cstp/english/quantum/index.html},
    urldate = {2026-01-07}
}

@online{korea,
  title        = {National Quantum Strategy},
  organization = {Quantum in Korea},
  year         = {2023},
  url          = {https://quantuminkorea.org/national-quantum-strategy/},
    urldate = {2026-01-07}
}

@online{iran,
  title        = {
Iran to Establish First National Quantum Communication and Atomic Clock Labs
},
  organization = {Quantum Insider},
  year         = {2025},
  url          = {https://thequantuminsider.com/2025/10/30/iran-to-establish-first-national-quantum-communication-and-atomic-clock-labs/},
    urldate = {2026-01-07}
}

@online{turkey,
  title        = {Erdogan unveils 2030 Industry and Technology Strategy: A vision for technological independence},
  organization = {TRT World},
  year         = {2025},
  url          = {https://www.trtworld.com/article/67385992a0a7},
    urldate = {2026-01-07}
}

@online{brazil,
  title        = {FAPESP pretende impulsionar o desenvolvimento de tecnologias quânticas no Brasil},
  organization = {Agência FAPESP},
  year         = {2024},
  url          = {https://agencia.fapesp.br/fapesp-pretende-impulsionar-o-desenvolvimento-de-tecnologias-quanticas-no-brasil/53594},
    urldate = {2026-01-07}
}

@online{peru,
  title        = {Tecnologías cuánticas},
  organization = {Government of Peru},
  url          = {https://www.gob.pe/quantumperu},
    urldate = {2026-01-07}
}

@online{LeonardoSaudi,
  title        = {
Leonardo signs MoU in the Kingdom of Saudi Arabia to expand collaboration in the aerospace and defence sector
},
  organization = {Leonardo},
  url          = {https://www.leonardo.com/en/press-release-detail/-/detail/26-01-2025-leonardo-signs-mou-in-the-kingdom-of-saudi-arabia-to-expand-collaboration-in-the-aerospace-and-defence-sector},
  year         = {2025},
    urldate = {2026-01-07}
}

@online{LeonardoEgypt,
  title        = {
Egypt close to completing \$3bn arms deal with Italy
},
  organization = {Middle East Eye},
  url          = {https://www.middleeasteye.net/news/egypt-italy-arms-deal-close-completing},
  year         = {2022},
    urldate = {2026-01-07}
}

@online{euDualUse2024,
  title        = {WHITE PAPER -
On options for enhancing support for research and development involving technologies
with dual-use potentia},
  organization = {European Commission},
  year         = {2024},
  url          = {https://research-and-innovation.ec.europa.eu/document/download/7ae11ca9-9ff5-4d0f-a097-86a719ed6892_en},
    urldate = {2026-01-07}
}

@online{santopinto2025,
  title        = {The REARM Europe Plan: Squaring the Circle Between Integration and National Sovereignty},
  author    = {Federico Santopinto},
organization = {
Institut de Relations Internationales et Stratégiques (IRIS)
},
  year         = {2025},
  url          = {https://www.iris-france.org/en/the-rearm-europe-plan-squaring-the-circle-between-integration-and-national-sovereignty/},
    urldate = {2026-01-07}
}

@online{consilium2025,
  title        = {EU defence in numbers},
  organization = {Council of the European Union},
  url          = {https://www.consilium.europa.eu/en/policies/defence-numbers/},
    urldate = {2026-01-07}
}

@online{sipri2025milex,
  title        = {Military Expenditure Database },
  organization = {Stockholm International Peace Research Institute (SIPRI)},
  url          = {https://doi.org/10.55163/CQGC9685},
    urldate = {2026-01-07}
}

@online{defensepost2025,
  organization       = {The Defense Post},
  title        = {EU Defense Spending Hits \$443B in 2025},
  year         = {2025},
  url          = {https://thedefensepost.com/2025/09/02/eu-defense-spending-record/},
    urldate = {2026-01-07}
}

@online{eurasiantimes2025,
  organization       = {The Eurasian Times},
author = {Sumit Ahlawat},
title        = {EU Beats China, Russia In Defense Spending; Significantly Behind The US In Fighter Jets, Ahead In MBTs, IFVs},
  year         = {2025},
  url          = {https://www.eurasiantimes.com/outpaced-in-fighter-jets-eu-surpasses-china-russia/},
    urldate = {2026-01-07}
}

@online{reuters2025nato,
  author       = {Lili Bayer and Andrew Gray},
organization = {Reuters},
  title        = {What is NATO's New 5\% Defence Spending Target?},
  year         = {2025},
  url          = {https://www.reuters.com/business/aerospace-defense/what-is-natos-new-5-defence-spending-target-2025-06-23/},
    urldate = {2026-01-07}
}

@online{reuters2025italy,
  organization       = {Reuters},
  title        = {Italy Says It Needs at Least 10 Years to Raise Defence Spending},
  year         = {2025},
  url          = {https://www.reuters.com/business/aerospace-defense/italy-says-it-needs-least-10-years-raise-defence-spending-2025-06-12/},
    urldate = {2026-01-07}
}

@online{europarl2025,
  title        = {ReArm Europe Plan/Readiness 2030},
  organization = {European Parliament},
  year         = {2025},
  url          = {https://www.europarl.europa.eu/thinktank/en/document/EPRS_BRI(2025)769566},
    urldate = {2026-01-07}
}

@article{krelina2021quantum,
  title={Quantum technology for military applications},
  author={Krelina, Michal},
  journal={EPJ Quantum Technology},
  volume={8},
  pages={24},
  year={2021},
url = {https://doi.org/10.1140/epjqt/s40507-021-00113-y},
  publisher={Springer Berlin Heidelberg}
}

@online{atarc2025,
author = {ATARC Quantum Working Group},
  organization = {Advanced Technology Academic Research Center (ATARC)},
  title        = {White Paper: Applied Quantum Computing for Today's Military},
  year         = {2021},
  url          = {https://atarc.org/project/white-paper-applied-quantum-computing-for-todays-military/},
    urldate = {2026-01-07}
}

@online{carnegie2019,
  author       = {Steve Feldstein},
  title        = {The Global Expansion of AI Surveillance},
  organization = {Carnegie Endowment for International Peace},
  year         = {2019},
  url          = {https://carnegieendowment.org/research/2019/09/the-global-expansion-of-ai-surveillance?lang=en},
    urldate = {2026-01-07}
}

@online{nato2025topics,
  title        = {Emerging and disruptive technologies},
  organization = {NATO},
  year         = {2025},
  url          = {https://www.nato.int/en/what-we-do/deterrence-and-defence/emerging-and-disruptive-technologies},
    urldate = {2026-01-07}
}

@online{nato2025texts,
  title        = {Summary of NATO’s Quantum Technologies Strategy},
  organization = {NATO},
  year         = {2024},
  url          = {https://www.nato.int/cps/en/natohq/official_texts_221777.htm},
    urldate = {2026-01-07}
}

@online{eu2025quantum,
  organization = {European Commission},
  title        = {Quantum Technologies},
  year         = {2025},
  url          = {https://defence-industry-space.ec.europa.eu/eu-space/research-development-and-innovation/quantum-technologies_en},
    urldate = {2026-01-07}
}

@online{ansa2025,
  organization = {ANSA},
  title        = {If You Want Peace, Prepare for War, Says PM Citing Romans},
  year         = {2025},
  url          = {https://www.ansa.it/amp/english/newswire/english_service/2025/06/24/if-you-want-peace-prepare-for-war-says-pm-citing-romans-4_4b34b41e-5f1c-40bd-9e2f-c960331549fe.html},
    urldate = {2026-01-07}
}

@online{pbs2025,
author = {Pan Pylas},
  title        = {UK's Starmer Urges Putin to Agree to Ukraine Ceasefire},
  organization = {PBS News},
  year         = {2025},
  url          = {https://www.pbs.org/newshour/world/uks-starmer-urges-putin-to-prove-he-is-serious-about-peace-by-agreeing-to-ukraine-ceasefire},
    urldate = {2026-01-07}
}

@online{devcom2025,
  organization = {US Army 
Combat Capabilities Development Command
 (DEVCOM)},
  title        = {Partnerships},
  url          = {https://devcom.army.mil/partner-with-us},
    urldate = {2026-01-07}
}

@online{fapesp2025,
author = {Rodrigo de Oliveira Andrade},
  organization = {Revista Pesquisa FAPESP},
  title        = {US Armed Forces Support Basic Research in Brazil},
  year         = {2022},
  url          = {https://revistapesquisa.fapesp.br/en/us-armed-forces-support-basic-research-in-brazil/},
    urldate = {2026-01-07}
}

@techreport{sipri2025fact,
  author       = {Xiao Liang and Nan Tian and Diego Lopes da Silva and Lorenzo Scarazzato and Zubaida A. Karim and Jade Guiberteau Ricard
},
  title        = {Trends in World Military Expenditure, 2024},
  year         = {2025},
  institution  = {Stockholm International Peace Research Institute (SIPRI)},
  type         = {SIPRI Fact Sheet},
  url          = {https://sipri.org/publications/2025/sipri-fact-sheets/trends-world-military-expenditure-2024}
}

@online{sipri2025press,
  title        = {Unprecedented Rise in Global Military Expenditure},
  organization = {Stockholm International Peace Research Institute (SIPRI)},
  year         = {2025},
  url          = {https://www.sipri.org/media/press-release/2025/unprecedented-rise-global-military-expenditure-european-and-middle-east-spending-surges},
    urldate = {2026-01-07}
}

@online{undp2025,
  title        = {Record Military Spending Threatens Global Peace and Development, New UN Report Warns},
  organization = {United Nations Development Programme (UNDP)},
  year         = {2025},
  url          = {https://www.undp.org/press-releases/record-military-spending-threatens-global-peace-and-development-new-un-report-warns},
    urldate = {2026-01-07}
}

@online{iiss2025,
author = {
    Fenella McGerty and Karl Dewey},
  title        = {Global Defence Spending Soars to New High},
  organization = {The International Institute for Strategic Studies (IISS)},
  year         = {2025},
  url          = {https://www.iiss.org/online-analysis/military-balance/2025/02/global-defence-spending-soars-to-new-high/},
    urldate = {2026-01-07}
}

@online{reuters2025allnato,
  organization       = {Reuters},
  title        = {All NATO Members Hit Old Spending Target; Only Three Meet New Goal},
  year         = {2025},
  url          = {https://www.reuters.com/business/aerospace-defense/all-nato-members-hit-old-spending-target-only-three-meet-new-goal-2025-08-27/},
    urldate = {2026-01-07}
}

@online{quantumzeitgeist2025,
  title        = {Russia Unveils 50-Qubit Quantum Computer Prototype},
  organization = {Quantum Zeitgeist},
  year         = {2025},
  url          = {https://quantumzeitgeist.com/russia-unveils-50-qubit-quantum-computer-prototype/},
    urldate = {2026-01-07}
}

@online{itif2024,
  author       = {Hodan Omaar and Martin Makaryan},
organization = {Information Technology \% Innovation Foundation (ITIF)},
  title        = {How Innovative is China in Quantum?},
  year         = {2024},
  url          = {https://itif.org/publications/2024/09/09/how-innovative-is-china-in-quantum/},
    urldate = {2026-01-07}
}

@online{gov2020,
  title        = {China to include quantum technology in its 14th Five-Year Plan},
  organization = {The State Council of the People's Republic of China},
  year         = {2020},
  url          = {https://english.www.gov.cn/news/videos/202010/22/content_WS5f90e700c6d0f7257693e3fe.html},
    urldate = {2026-01-07}
}

@online{quantuminsider2025,
  title        = {Quantum, AI, and Space Anchor Pentagon’s Deep-Tech Convergence Strategy},
author = {Matt Swayne},
  organization = {The Quantum Insider},
  year         = {2025},
  url          = {https://thequantuminsider.com/2025/07/03/quantum-ai-and-space-anchor-pentagons-deep-tech-convergence-strategy/},
    urldate = {2026-01-07}
}

@online{dod2025,
  title        = {Defense Budget Materials - FY2025},
  organization = {US Department of Defense},
  year         = {2025},
  url          = {https://comptroller.war.gov/Budget-Materials/Budget2025/},
    urldate = {2026-01-07}
}

@online{dod2024,
  title        = {Department of Defense
Fiscal Year (FY) 2025 Budget Estimates},
  organization = {US Department of Defense},
  year         = {2024},
  url          = {https://comptroller.defense.gov/Portals/45/Documents/defbudget/FY2025/budget_justification/pdfs/03_RDT_and_E/RDTE_OSD_PB_2025.pdf},
    urldate = {2026-01-07}
}

@online{spiegel2023,
  author       = {Der Spiegel},
  title        = {Hochschulen müssen aufhören, sich hinter der Zivilklausel zu verstecken},
author = {Thomas Hofmann},
  year         = {2025},
  url          = {https://www.spiegel.de/politik/deutschland/aufruestung-universitaeten-sollten-die-militaerische-forschung-verstaerken-meinung-a-6b9f95e1-cb78-4193-8668-a35799155119},
    urldate = {2026-01-07}
}

@online{gao2023,
  organization = {US Government Accountability Office},
  title        = {Federal Research and Development},
  year         = {2022},
  url          = {https://www.gao.gov/products/gao-23-105396},
    urldate = {2026-01-07}
}

@online{economicstrategy2025,
  organization = {Aspen Economic Strategy Group},
  title        = {Seven Recent Developments in US Science Funding},
  year         = {2023},
  url          = {https://www.economicstrategygroup.org/publication/seven-recent-developments/},
    urldate = {2026-01-07}
}

@online{spie2025,
  organization = {SPIE},
  title        = {Quantum Technologies for Defence and Security Conference},
  year         = {2025},
  url          = {https://spie.org/ESI25D/conferencedetails/quantum-technologies-defence-and-security},
    urldate = {2026-01-07}
}

@online{ilmanifesto2025,
  title        = {Leonardo e la militarizzazione della ricerca accademica},
author = {Paola Rivetti},
  organization = {Il Manifesto},
  year         = {2024},
  url          = {https://ilmanifesto.it/leonardo-e-la-militarizzazione-della-ricerca-accademica},
    urldate = {2026-01-07}
}

@online{leonardo2023,
  organization = {Leonardo},
  title        = {Leonardo Signs Agreements with Israeli Innovation Authority and Ramot Tel Aviv University},
  year         = {2023},
  url          = {https://www.leonardo.com/en/press-release-detail/-/detail/03-02-2023-leonardo-signs-two-agreements-with-israeli-innovation-authority-and-ramot-tel-aviv-university-in-the-field-of-innovation},
    urldate = {2026-01-07}
}

@online{ohchr2025,
  organization = {Office of the United Nations High Commissioner for Human Rights (OHCHR)},
  title        = {Israel Has Committed Genocide in Gaza Strip, UN Commission Finds},
  year         = {2025},
  url          = {https://www.ohchr.org/en/press-releases/2025/09/israel-has-committed-genocide-gaza-strip-un-commission-finds},
    urldate = {2026-01-07}
}

@online{altreconomia2025,
author= {Duccio Facchini},
  title        = {Leonardo ammette l’export di armi in Israele e fa cadere la maschera del governo},
  organization = {Altrəconomia},
  year         = {2025},
  url          = {https://altreconomia.it/leonardo-ammette-lexport-di-armi-in-israele-e-fa-cadere-la-maschera-del-governo/},
    urldate = {2026-01-07}
}

\end{document}